\documentclass[sigconf,screen]{acmart}

\setcopyright{none}

\usepackage{tabularx}
\usepackage{array}
\usepackage{threeparttable}
\usepackage{booktabs}
\usepackage{subfiles}
\usepackage{multirow}
\usepackage{tikz}
\usetikzlibrary{arrows.meta,calc}
\usepackage[table]{xcolor}
\usepackage{dblfloatfix} 
\newif\ifcolouredlinks
\colouredlinkstrue        
\makeatletter
\renewenvironment{quote}
  {\list{}{\leftmargin=1em \rightmargin=1em}%
   \item\relax}
  {\endlist}
\makeatother

\definecolor{citeBlue}{HTML}{0B63C9}   
\definecolor{sessPurple}{HTML}{AA00FF} 

\AtBeginDocument{%
  \ifcolouredlinks
    \hypersetup{
      colorlinks=true,
      linkcolor=black,     
      citecolor=citeBlue,  
      urlcolor=citeBlue    
    }%
  \else
    \hypersetup{colorlinks=false}%
  \fi
}

\newcommand{\SessNumLower}{0.0ex}     
\newcommand{\SessSSize}{\normalsize}   
\newcommand{\SessNumSize}{\small} 

\newcommand{\SessS}{\SessSSize S}
\newcommand{\SessNum}[1]{\raisebox{-\SessNumLower}[0ex][0ex]{\SessNumSize #1}}

\newcommand{\SessMark}[1]{%
  \mbox{\hypertarget{sess:#1}{\SessS\SessNum{#1}}}%
}

\newcommand{\SessTokStart}[1]{%
  \begingroup
    \hypersetup{linkcolor=sessPurple}%
    \mbox{\hyperlink{sess:#1}{\SessS\SessNum{#1}}}%
  \endgroup
}

\newcommand{\SessTokNum}[1]{%
  \begingroup
    \hypersetup{linkcolor=sessPurple}%
    \mbox{\hyperlink{sess:#1}{\SessNum{#1}}}%
  \endgroup
}

\ExplSyntaxOn
\tl_new:N   \l__jl_out_tl
\seq_new:N  \l__jl_seq
\int_new:N  \l__jl_start_int
\int_new:N  \l__jl_prev_int
\bool_new:N \l__jl_first_bool

\cs_new_protected:Npn \__jl_sess_flush:
  {
    \bool_if:NF \l__jl_first_bool { \tl_put_right:Nn \l__jl_out_tl { ;  } }
    \int_compare:nNnTF { \l__jl_start_int } = { \l__jl_prev_int }
      { 
        \tl_put_right:Nx \l__jl_out_tl
          { \exp_not:N \SessTokStart { \int_use:N \l__jl_start_int } }
      }
      { 
        \tl_put_right:Nx \l__jl_out_tl
          {
            \exp_not:N \SessTokStart { \int_use:N \l__jl_start_int }
            -- 
            \exp_not:N \SessTokNum   { \int_use:N \l__jl_prev_int  }
          }
      }
    \bool_set_false:N \l__jl_first_bool
  }

\cs_new_protected:Npn \jl_sess_list:n #1
  {
    \seq_set_split:Nnn \l__jl_seq { , } { #1 }
    \seq_sort:Nn \l__jl_seq
      {
        \int_compare:nNnTF { \int_eval:n { ##1 } } < { \int_eval:n { ##2 } }
          { \sort_return_same: } { \sort_return_swapped: }
      }

    \tl_clear:N \l__jl_out_tl
    \bool_set_true:N \l__jl_first_bool
    \int_zero:N \l__jl_start_int
    \int_zero:N \l__jl_prev_int

    \seq_map_inline:Nn \l__jl_seq
      {
        \tl_set:Nn \l_tmpa_tl { ##1 }
        \tl_trim_spaces:N \l_tmpa_tl
        \int_set:Nn \l_tmpa_int { \tl_use:N \l_tmpa_tl }

        \int_compare:nNnTF { \l__jl_prev_int } = { 0 }
          { 
            \int_set_eq:NN \l__jl_start_int \l_tmpa_int
            \int_set_eq:NN \l__jl_prev_int  \l_tmpa_int
          }
          {
            \int_compare:nNnTF { \l_tmpa_int } = { \l__jl_prev_int }
              { }
              {
                \int_compare:nNnTF { \l_tmpa_int } = { \l__jl_prev_int + 1 }
                  { \int_set_eq:NN \l__jl_prev_int \l_tmpa_int }
                  {
                    \__jl_sess_flush:
                    \int_set_eq:NN \l__jl_start_int \l_tmpa_int
                    \int_set_eq:NN \l__jl_prev_int  \l_tmpa_int
                  }
              }
          }
      }
    \int_compare:nNnT { \l__jl_prev_int } > { 0 } { \__jl_sess_flush: }
    \tl_use:N \l__jl_out_tl
  }

\NewDocumentCommand{\session}{m}{[\jl_sess_list:n{#1}]}

\ExplSyntaxOff

\renewcommand{\SessMark}[1]{%
  \mbox{\hypertarget{sess:#1}{\SessS\SessNum{#1}}}%
}

\usetikzlibrary{arrows.meta} 

\usetikzlibrary{calc}

\definecolor{naturalColor}{RGB}{8,230,0}
\definecolor{LatentColor}{RGB}{69,150,66}
\definecolor{externalColor}{RGB}{164,224,34}
\definecolor{featureColor}{RGB}{71,179,255}
\definecolor{inferenceColor}{RGB}{56,112,232}
\definecolor{applicationColor}{RGB}{107,47,247}
\definecolor{outcomecolor}{RGB}{66, 245, 173}
\definecolor{outcomeColor}{RGB}{66, 245, 173}

\title{Monitoring and Observability of Machine Learning Systems: Current Practices and Gaps}

\author{Joran Leest}
\affiliation{
  \institution{Vrije Universiteit Amsterdam}
  \country{The Netherlands}
}
\email{j.g.leest@vu.nl}

\author{Ilias Gerostathopoulos}
\affiliation{
  \institution{Vrije Universiteit Amsterdam}
  \country{The Netherlands}
}
\email{i.g.gerostathopoulos@vu.nl}

\author{Patricia Lago}
\affiliation{
  \institution{Vrije Universiteit Amsterdam}
  \country{The Netherlands}
}
\email{p.lago@vu}

\author{Claudia Raibulet}
\affiliation{
  \institution{Universita' degli Studi di Milano-Bicocca}
  \country{Italy}
}
\email{claudia.raibulet@unimib.it}

\begin{document}

\begin{abstract}
    Production machine learning (ML) systems fail silently -- not with crashes, but through wrong decisions. While observability is recognized as critical for ML operations, there is a lack empirical evidence of what practitioners actually capture. This study presents empirical results on ML observability in practice through seven focus group sessions in several domains. We catalog the information practitioners systematically capture across ML systems and their environment and map how they use it to validate models, detect and diagnose faults, and explain observed degradations. Finally, we identify gaps in current practice and outline implications for tooling design and research to establish ML observability practices.
\end{abstract}
\maketitle

\section{Introduction}\label{intro}

Software monitoring has matured through decades of practice and research, enabling engineers to track system health, validate correctness, and detect degradations using established reliability practices \cite{beyer2018site}. At its core, monitoring is enabled through observability \cite{cantrill2006hidden,lyu2007software,mahida2023enhancing,niedermaier2019observability,usman2022survey} -- the ability to measure and understand a system's behavior during operation \cite{gorton2023observability,usman2022survey}. Practically, this means capturing the system context as logs, metrics, and traces \cite{gatev2021observability}. As systems increasingly incorporate data pipelines and warehouses as core infrastructure, the notion of observability has extended its scope to include data quality metrics and lineage, enabling monitoring of these components \cite{petrella2023fundamentals,breck2019data,swami2020data}.

Today, with the pervasive use of machine learning (ML), monitoring requires observability that extends beyond code and data to also capture learned behaviors -- information about how models decide which customers receive loans, which content gets promoted, or which transactions trigger fraud alerts. Unlike traditional software, when ML systems fail, they fail silently -- not with crashes or error logs, but by making wrong decisions \cite{bennett2022silent,bungert2023understanding,sculley2015hidden,shankars2022towards}.

These silent failures point to the challenge of monitoring ML systems, which requires new observability approaches that existing frameworks and tools do not yet support, as we discuss below.

\textbf{Monitoring ML systems requires distinguishing real failures from benign changes, which proves to be inherently ambiguous.} When faults occur, they typically surface as distributional changes -- changes in the statistical properties of the data that can degrade model performance, also known as data drift. \cite{gama2014survey,webb2016characterizing}. These changes arise from multiple sources \cite{shankars2024we,huyen2022}: internal faults such as failed processing jobs or corrupted data pipelines, and external factors such as trends, seasonality, and events. Neither their source nor magnitude signals the need for intervention. Internal errors and data quality flags range from benign to critical; external changes span from natural fluctuations to fundamental changes in the data-generating process. Moreover, teams often lack timely label feedback to assess whether data drift actually degrades model performance \cite{huyen2022,ginartaa2022mldemon,shankars2024we,zheng2019labelless,shergadwalamn2022a}, leaving practitioners to observe drifts without reliably determining their cause or impact.

\textbf{This ambiguity arises because ML systems simultaneously function as computational systems and as models of external processes, requiring observability that captures both views.} Synthesizing the literature, the authors of \cite{leest2025tea} conclude that effective monitoring indeed requires capturing information across both views of the system context: the ML system (dependency graphs and metadata showing how data, models, and applications interact) and the external environment (tracking exogenous drivers like sales campaigns or trends that explain distribution changes). They develop a descriptive model classifying the system context information needed for ML observability.

\textbf{Yet existing tools do not capture this unified system context.} Tools provide capabilities for data lineage \cite{datadog,montecarlo}, data validation \cite{greatexpectations, deequ}, and drift detection \cite{arize, fiddler}, yet ML teams rarely receive actionable signals \cite{shergadwalamn2022a, shankars2024we, protschky2025gets}. We argue that the problem is fragmented context. Tools capture pieces of information in isolation rather than connecting them. When alerts are fired, practitioners must correlate relevant signals manually -- matching error flags across systems, contacting upstream teams, and consulting domain experts \cite{ebreck2017the, shergadwalamn2022a, shankars2024we, protschky2025gets}. A recent survey on ML operations (MLOps) confirms the tooling gap: monitoring and observability rank as the top challenge, with 77\% of the respondents reporting either no monitoring tools or custom-built solutions \cite{EthicalInstituteStateOfProdML2024}.

\textbf{We lack empirical evidence on how practitioners actually achieve ML observability in practice.} Despite growing interest, there is not yet a commonly accepted framework for practice \cite{shankars2022towards,nogare2024machine,palumbo2023observability}. The descriptive model proposed in \cite{leest2025tea} offers initial understanding, but frameworks derived from literature may diverge from actual practice. This motivates our research question:

\begin{quote}
\textit{What information of the system context do practitioners use when monitoring an ML system?}
\end{quote}

We present an empirical study of ML observability across organisations and domains. Guided by an existing descriptive model~\cite{leest2025tea} (Section \ref{background}), we carried out focus groups and analyzed their results with framework analysis (Section \ref{study_design}). We describe what teams capture about system context, how they use it in monitoring, and where gaps and opportunities remain (Section \ref{results}). Our primary contribution is the first empirical characterization of ML observability. As a secondary contribution, we assess the model’s fit to practice and propose refinements to the model (Section~\ref{discussion}).

\section{Background: Descriptive Model of System Context Information}\label{background}

We use the descriptive model from \cite{leest2025tea} to organize our findings, characterizing information relevant to ML monitoring by \textit{system context} and \textit{information type}. This enables us to catalog what practitioners capture -- from core signals like feature and prediction statistics to broader contextual information.

\subsection{System Context}

The \textit{system} dimension specifies the context of which information is captured. Whereas traditional observability centers on computational artefacts (e.g., code or data artefacts), ML systems require a broader perspective that captures both the technical ML system itself and the natural environment that generates the data it processes.
In \cite{leest2025tea}, the authors distinguish two views of system context: the natural \textit{environment} and the technical \textit{ML system}.

The \textit{environment} represents the data-generating processes external to the ML system -- the view a domain expert adopts when diagnosing faults in the model (e.g., “did an external event cause the performance drop?”). It includes three subsystems:
\begin{itemize}
    \item \textit{Reference domain} -- the real-world process the model aims to capture, expressed as unobserved random variables over \textit{features} (e.g., customer age) and a \emph{target}. The target is the predicted variable, a domain property invariant to the ML system’s operation, capturing either a realised event (e.g., a fraudulent transaction) or a propensity towards a future event (e.g., a defaulted loan).
    \item \textit{Exogenous influences} -- observable \textit{exogenous variables} that drive the modeled domain (e.g., weather conditions);
    \item \textit{Latent influences} -- unobserved \textit{latent variables} that can only be inferred from statistical patterns (e.g., customer trends).
\end{itemize}

The \textit{ML system} covers the managed technical artefacts relevant to running the model -- the engineer’s view when interpreting an observed drift (e.g., “was it a failed job, corrupted data, or an infrastructure failure?”). It includes the following subsystems:
\begin{itemize}
    \item \textit{Processing pipeline} -- \textit{data transformations} from \textit{source tables} to \textit{feature tables} (e.g., clicks and sessions aggregated into a weekly activity feature);
    \item \textit{Inference pipeline} -- the serving layer where an ML model consumes features to produce predictions (e.g., a recommendation model producing ranked product preferences);
    \item \textit{Application} -- downstream decision policies converting predictions into actions (e.g., selecting a personalized promotion from ranked preferences).
\end{itemize}

These subsystems are causally ordered: exogenous and latent influences drive the reference domain; behavior is observed and engineered into features; the model maps features to predictions; and application policies turn predictions into actions.

\subsection{Information Type} \label{information_types}

The type of information that can be captured of the system context varies by \textit{aspect} (states, structures, or properties) asserted about that element, and the format in which this is \textit{represented} (formal or informal) \cite{leest2025tea}. For brevity, we list the unique combinations between these dimensions observed in our study as \textit{information types}:

\begin{itemize}
    \item \textit{Attributes} -- numerical measures of the conditions under which a system element operates (e.g., inference latency).
    \item \textit{Metadata} -- semi-structured descriptions of a system element’s conditions (e.g., a table schema).
    \item \textit{Metrics} -- numerical measures that evaluate a system element (e.g., distribution divergence).
    \item \textit{Identifiers} -- numerical key-value measures that identify records as belonging to an entity (e.g., customer geography).
    \item \textit{Slices} -- logical definitions of entities to partition records (e.g., adult customers are aged 18-65).
    \item \textit{Graphs} -- graphical representations of relations between system elements (e.g., causal diagram of feature variables).
    \item \textit{Assertions} -- logical predicates that encode the properties of a system element (e.g., customer age is in range [0, 100]).
\end{itemize}

\section{Study Design and Execution}\label{study_design}

To address our main research question (Section \ref{intro}), we conducted focus group sessions exploring two sub-questions: \textit{``What information of the system context do practitioners capture and use in monitoring their ML system?''} (RQ1), and \textit{``What information is used, but not systematically captured (or only informally managed)?''} (RQ2). We elicit current ML observability practices -- the system information practitioners systematically capture and rely on, organizing items according to the descriptive model reused from \cite{leest2025tea}, and identify gaps in the captured information.

\subsection{Participant Recruiting}

We recruited participants through LinkedIn\footnote{https://www.linkedin.com/feed/}, Slack MLOps communities\footnote{https://mlops.community/}, and our professional network. The recruitment message targeted anyone involved in monitoring ML systems, explicitly welcoming diverse roles including but not limited to data scientists, ML/data/software engineers, platform engineers, and business stakeholders, mitigating role representation bias by capturing holistic perspectives rather than preconceived notions about roles. We mitigated selection bias by recruiting through multiple channels and including participants at any experience level.

\subsection{Sessions}
We conducted seven focus group sessions in six organizations that span the banking, financial technology, industrial technology, and e-commerce domains. Sessions lasted 54-82 minutes, with 4-9 participants in each with a total of 37 participants.

Participants represented four self-declared roles: (1) \textit{data scientists} (including ML scientists) focusing on model development; (2) \textit{ML engineers} (including MLOps engineers) handling pipeline implementations and production operations; (3) \textit{platform engineers} (including infrastructure engineers) managing ML tooling and infrastructure; and (4) \textit{business stakeholders} (project managers, product owners) providing organizational context.


We summarize the sessions information in Table \ref{tab:focus_group_sessions}. Notice that sessions \session{1} and \session{2} were from the same organization but involved different teams, non-overlapping participants, and distinct sub-domains. We therefore treat these as separate sessions.

\begin{table}[htbp]
\centering
\caption{Anonymised descriptions of focus group sessions}
\label{tab:focus_group_sessions}
\begin{threeparttable}
\setlength{\tabcolsep}{4.4pt}
\renewcommand{\arraystretch}{1.1}
\begin{tabular}{@{}p{0.3cm}|p{1.9cm}|p{2.2cm}|p{1.2cm}|p{1.6cm}@{}}
\toprule
\textbf{ID} & \textbf{Domain} & \textbf{Participants}\tnote{$\dagger$} & \textbf{Org. Size} & \textbf{Maturity} \\
\midrule
\SessMark{1}\tnote{*} & Banking         & 5 MLE                     & Large  & Early \\
\SessMark{2}\tnote{*} & Banking         & 2 DS, 2 MLE               & Large  & Early \\
\SessMark{3}          & Financial Tech  & 4 DS, 2 MLE               & Medium & Advanced \\
\SessMark{4}          & Industrial Tech & 3 DS, 1 MLE               & Small  & Early \\
\SessMark{5}          & E-commerce      & 4 DS, 1 MLE               & Large  & Early \\
\SessMark{6}          & E-commerce      & 1 DS, 3 PE                & Medium & Intermediate \\
\SessMark{7}          & Banking         & 3 DS, 4 MLE, 2 BS         & Large  & Intermediate \\
\bottomrule
\end{tabular}
\begin{tablenotes}
\vspace{0.5em}
\footnotesize
\item[*] S1 and S2 represent separate sessions conducted with the same organization
\item[$\dagger$] DS = Data Scientist; MLE = ML Eng.; PE = Platform Eng.; BS = Business Stakeholder
\end{tablenotes}
\end{threeparttable}
\end{table}

We defined organization size by employee count: small (<1,000), medium (1,000-10,000), and large (>10,000). We assessed monitoring maturity based on participants' self-reported experience and extent of tooling in place. \textit{Early stage maturity} indicates primary reliance on manual processes with basic tracking that teams rarely act upon. \textit{Intermediate stage maturity} involves active tracking and alerting but maintains largely manual processes with significant coverage gaps. \textit{Advanced stage maturity} includes organizations with established monitoring platforms, though may still rely on manual processes.

We mitigated threats to generalizability by continuing recruitment until diversity was reached across roles, domains, and maturity levels, as shown in Table \ref{tab:focus_group_sessions}.

\subsection{Procedure}

Sessions used a printed poster displaying the six subsystems from the descriptive model reused from \cite{leest2025tea} (reference domain, exogenous influences, and latent influences for the external environment; processing pipeline, inference pipeline, and application for the ML system). The facilitators used post-it notes to document \textit{general information} of the case, \textit{where} incidents occurred, and \textit{what} information was consulted throughout the session -- using brief in-session member checks to verify interpretation \cite{candela2019exploring}.

Each session followed four phases. \textbf{First}, we elicited monitoring challenges and priorities to understand what practitioners find problematic. \textbf{Second}, we mapped participants' ML case onto the poster -- e.g., the complexity of their data processing, the model domain, their application -- for a shared understanding. \textbf{Third}, we elicited incident scenarios' deep-dives where participants walked through specific monitoring incidents, describing how issues surfaced, how they investigated, and how they diagnosed root causes. \textbf{Finally}, we conducted a wrap-up with member checking.

We mitigated researcher bias in how we introduced the descriptive model. We explained the six subsystems shown on the poster as a structure for discussion -- using it to organize post-it notes consolidating their case -- but did not explain the types of information that could be captured. We elicited information through neutral questions centered around monitoring practice -- e.g., ``how did the incident surface?'', ``how did you find the root-cause?'' -- rather than directly prompting information usage to avoid bias. This approach enabled validation of the reused framework's dimensions: The system context dimension was validated through participants recognizing their system in the provided structure, while the information type dimension was validated through inductive coding of naturally described information. We mitigated construct validity threats by including an open field on the poster for unmappable information, and ensured balanced role representation through round-robins and inclusive prompting of participants.

\subsection{Data Collection and Analysis}\label{analysis}

We recorded and transcribed all focus group sessions. The transcripts served as our primary data source for analysis, with collected post-it notes as supplementary materials.

Again, to address RQ1, we employed framework analysis \cite{goldsmith2021using}, using the descriptive model from \cite{leest2025tea} as our a priori analytical frame. We deductively coded \emph{information items} -- self-contained statements of explicitly captured information relevant to ML monitoring. While coding was performed at the full granularity of the model, we report simplified information-type categories (Section~\ref{information_types}) for clarity. Applying the model as our coding frame provides evidence for its construct validity and face validity in operational settings. To mitigate the risk of forcing data into predefined categories, we treated unmappable content inductively. Recurrent unmappable content indicated systematic gaps in the original model; we formalized these as extensions to the descriptive model (Section~\ref{suggestions}) and integrated them into our results.

For RQ2, we inductively coded \emph{uncaptured information items} -- items that participants reported using or needing in practice but that are not systematically captured or available to the monitoring workflow (e.g., inaccessible sources, tacit knowledge).

We annotated each information item (RQ1, RQ2) with its partici-pant-described monitoring use and grouped items by similar uses (e.g., fault detection, localization) into six monitoring activities. 


\section{Results}\label{results}

\newcommand{\dashedbox}{\tikz[baseline=0.35ex] \draw[fill=none, line width=0.3pt, dashed, rounded corners=2pt, line cap=butt,
      dash pattern=on 2pt off 1.2pt,
      minimum width=20pt,
      minimum height=5pt] (0,0) rectangle (0.5,0.25);}
\newcommand{\solidbox}{\tikz[baseline=0.35ex] \draw[fill=none, line width=0.5pt, rounded corners=2pt] (0,0) rectangle (0.5,0.25);}

    \newcommand{\FuncSymbol}{\tikz[baseline=-0.6ex] \draw[-latex, line width=0.8pt, black] (-0.1,0) -- (0.4,0);}

\newcommand{\VarSymbol}[1]{%
  \tikz[baseline=-0.6ex] \fill[#1, draw=none] (0,0) circle[radius=0.1cm];%
}

\newcommand{\subsystemspacing}{-0.75cm}  
\newcommand{\boxopacity}{0.3}           
\newcommand{\boxlinewidth}{0.6pt}         
\newcommand{\roundedcorners}{4pt}       

\tikzset{
  subsystem arrow/.style={
    draw,                
    black,
    line width=0.1pt,
    -{Triangle[length=1.3mm, width=1mm]} 
  }
}

\newlength{\tablerowheight}
\setlength{\tablerowheight}{1.2\baselineskip}  

\newlength{\rowheightcm}
\setlength{\rowheightcm}{\tablerowheight}
\pgfmathsetmacro{\rowcm}{\rowheightcm/32}

\def\boxwidth{2}

\newcommand{\ExogenousBox}{%
\begin{tikzpicture}[baseline=(current bounding box.center), remember picture]
  \pgfmathsetmacro{\boxheight}{1.4*\rowcm}  
  \node[
    fill=externalColor,
    draw=black,
    dashed,                
      line cap=butt,  
    line width=\boxlinewidth,
    fill opacity=\boxopacity,
    rounded corners=\roundedcorners,
    minimum width=\boxwidth cm,
    minimum height=\boxheight cm,
    text=black,
  text opacity=1,    
    anchor=south west
  ] (exogenous) at (0,1.8) {};
  \node[anchor=center] at (\boxwidth/2,\boxheight/2+1.8) {\small Exogenous Var.};
\end{tikzpicture}%
}

\newcommand{\ReferenceBox}{%
\begin{tikzpicture}[baseline=(current bounding box.center), remember picture]
  \pgfmathsetmacro{\boxheight}{3.98*\rowcm}  
  \node[
    fill=naturalColor,
    draw=black,
    dashed,                
  line cap=butt,  
    line width=\boxlinewidth,
    fill opacity=\boxopacity,
    rounded corners=\roundedcorners,
    minimum width=\boxwidth cm,
    minimum height=\boxheight cm,
    anchor=south west
  ] (reference) at (0,0) {};
  \node[anchor=center] at (\boxwidth/2,\boxheight-1.3*\rowcm) {\small Feature Var.};
  \node[anchor=center] at (\boxwidth/2,\boxheight-3.3*\rowcm) {\small Target Var.};
\end{tikzpicture}%
}

\newcommand{\DataProcessingBox}{%
\begin{tikzpicture}[baseline=(current bounding box.center), remember picture]
  \pgfmathsetmacro{\boxheight}{8.76*\rowcm}  
  \node[
    fill=featureColor,
    draw=black,
    line width=\boxlinewidth,
    fill opacity=\boxopacity,
    rounded corners=\roundedcorners,
    minimum width=\boxwidth cm,
    minimum height=\boxheight cm,
    anchor=south west
  ] (dataprocessing) at (0,0) {};

  \node[anchor=center] at (\boxwidth/2,\boxheight-1.2*\rowcm) {\small Source Table};

  \node[anchor=center] at (\boxwidth/2,\boxheight-4.4*\rowcm) {\small Transformation};
  \node[anchor=center] at (\boxwidth/2,\boxheight-7.5*\rowcm) {\small Feature Table};
  \pgfmathsetmacro{\sepA}{\boxheight - 2.55*\rowcm}
  \pgfmathsetmacro{\sepB}{\boxheight - 4.90*\rowcm}

\end{tikzpicture}%
}

\newcommand{\ServingBox}{%
\begin{tikzpicture}[baseline=(current bounding box.center), remember picture]
  \pgfmathsetmacro{\boxheight}{5.02*\rowcm+0.006cm}  
  \node[
    fill=inferenceColor,
    draw=black,
    line width=\boxlinewidth,
    fill opacity=\boxopacity,
    rounded corners=\roundedcorners,
    minimum width=\boxwidth cm,
    minimum height=\boxheight cm,
    anchor=south west
  ] (serving) at (0,0) {};
  \node[anchor=center] at (\boxwidth/2,\boxheight-0.65*\rowcm) {\small Feature Vector\tnote{*}};
  \node[anchor=center] at (\boxwidth/2,\boxheight-2.7*\rowcm) {\small Model};
  \node[anchor=center] at (\boxwidth/2,\boxheight-4.68*\rowcm) {\small Predictions};
\end{tikzpicture}%
}

\newcommand{\ApplicationBox}{%
\begin{tikzpicture}[baseline=(current bounding box.center), remember picture]
  \pgfmathsetmacro{\boxheight}{2.86*\rowcm}  
  \node[
    fill=applicationColor,
    draw=black,
    line width=\boxlinewidth,
    fill opacity=\boxopacity,
    rounded corners=\roundedcorners,
    minimum width=\boxwidth cm,
    minimum height=\boxheight cm,
    anchor=south west
  ] (application) at (0,0) {};
  \node[anchor=center] at (\boxwidth/2,\boxheight-0.7*\rowcm) {\small Policy};
  \node[anchor=center] at (\boxwidth/2,\boxheight-2.1*\rowcm) {\small Actions};
\end{tikzpicture}%
}

\newcommand{\ResponseBox}{%
\begin{tikzpicture}[baseline=(current bounding box.center), remember picture]
  \pgfmathsetmacro{\boxheight}{1.4*\rowcm}  
  \node[
    fill=outcomecolor,
    draw=black,
    dashed,                
  line cap=butt,  
    line width=\boxlinewidth,
    fill opacity=\boxopacity,
    rounded corners=\roundedcorners,
    minimum width=\boxwidth cm,
    minimum height=\boxheight cm,
    text=black,
    text opacity=1,
    anchor=south west
  ] (response) at (0,1.8) {\small Outcome Var.\tnote{*}};
  \node[anchor=center] at (\boxwidth/2,0cm) {};
\end{tikzpicture}%
}

\begin{table*}[t]
\centering
\begin{threeparttable}  
\caption{Catalog of system information captured in ML monitoring practice, elicited from practitioner sessions. Columns show: \textit{System Elements} (the context of which information is captured), \textit{Information} (type of information captured), a \textbf{description} of each, concrete \textbf{examples}, a \textbf{count} of the number of unique sessions in which each appeared, and the corresponding session \textbf{ID}.}

\small
\setlength{\tabcolsep}{2.5pt}
\renewcommand{\arraystretch}{1.3}

\begin{tabular}{ @{} >{\centering\arraybackslash}p{2cm} >{\hspace{0.5em} \columncolor{gray!15}}l | >{\hspace{0.5em}}p{0.39\textwidth} | >{\hspace{0.5em}\itshape}p{0.20\textwidth} | c | p{0.088\textwidth} | }

\toprule
\multicolumn{1}{|l|}{\textbf{System Element}} & 
\multicolumn{1}{l|}{\textbf{Information}} &
\textbf{Description} & \textbf{Example} & \textbf{Count} & \textbf{ID} \\
\cmidrule{1-6}
\\[-0.4cm]

\cmidrule(l{7.2em}){1-6}
\\[-0.56cm]
\multirow{1}{*}{\ExogenousBox}
\\[-0.39cm]
 & Identifiers & Key-value mappings that partition data into subgroups & Device type: mobile/desktop & 3 & \session{5,6,3} \\
\\[-0.45cm]
\cmidrule(l{7.2em}){1-6}
\\[\dimexpr\subsystemspacing/2]

\\[\dimexpr\subsystemspacing/2]

\cmidrule(l{7.2em}){1-6}
\\[-0.56cm]
\multirow{3}{*}{\ReferenceBox}
\\[-0.385cm]
  & Slices & Expressions over features that partition data into subgroups. & Power users (sessions>20/week) & 1 & \session{7} \\
  & Assertions & Declarative domain constraints over feature variables. & User age in range [18, 100] & 2 & \session{6,2} \\
  \cmidrule{1-6}
  & Assertions & Declarative domain constraints over target variables. & Positive class rate: 4\%-6\% & 2 & \session{2,6} \\
  \\[-0.455cm]
\cmidrule(l{7.2em}){1-6}
\\[\dimexpr\subsystemspacing/2]
\\[\dimexpr\subsystemspacing/2]
\cmidrule(l{0.4em}){1-6}
\\[-0.56cm]
\multirow{7}{*}{\DataProcessingBox}
\\[-0.385cm]
& Metadata & Descriptive information of source data. & Table schema definitions & 2 & \session{2,3} \\
  & Metrics & Summary measures computed from source data. & Row counts, column nulls & 2 & \session{6,2} \\
  \cmidrule{1-6}
  & Attributes & Operational measures of a data operation's execution. & ETL job runtime & 2 & \session{7,2} \\
    & Metadata & Descriptive information of data operations. & Mapping config file & 2 & \session{6,7} \\
  & Graphs & Directed graph of data operations tracing the data lineage. & Data dependency DAG & 2 & \session{6,7} \\
  \cmidrule{1-6}
  & Attributes & Operational measures of stored feature columns. & Table freshness & 1 & \session{2} \\
  & Metrics & Summary measures over feature columns. & Column consistency & 3 & \session{6,1,2,3} \\
  \\[-0.46cm]
\cmidrule(l{0.4em}){1-6}
\\[-0.29cm]
\cmidrule(l{0.4em}){1-6}
\\[-0.56cm]
\multirow{3}{*}[0pt]{\ServingBox}
\\[-0.385cm]
  & Metrics & Summary measures from served input vectors per feature. & Null rates per feature & 5 & \session{6,7,2,5,3} \\
\cmidrule{1-6}
  & Attributes & Serving measures of the deployed model. & Inference latency & 3 & \session{4,5,7} \\
  & Metadata & Descriptive information of the model artefact. & Model card & 1 & \session{4} \\
\cmidrule{1-6}
  & Metrics & Summary measures computed from model outputs. & Predicted positive class rate & 6 & \session{5,6,7,1,2,3} \\
  \\[-0.46cm]
\cmidrule(l{0.4em}){1-6}
\\[\dimexpr\subsystemspacing/2]

\\[\dimexpr\subsystemspacing/2]
\cmidrule(l{0.4em}){1-6}
\\[-0.56cm]
\multirow{2}{*}{\ApplicationBox}
\\[-0.385cm]
  & Metadata & Descriptive information of the decision policy. & Decision config version & 1 & \session{3} \\
    \cmidrule{1-6}
  & Metrics & Summary measures of executed actions. & Intervention rate & 4 & \session{1,3,7,4} \\
  \\[-0.46cm]
\cmidrule(l{0.4em}){1-6}
\\[\dimexpr\subsystemspacing/2]

\\[\dimexpr\subsystemspacing/2]
\cmidrule(l{7.2em}){1-6}
\\[-0.56cm]
\multirow{2}{*}{\ResponseBox}
\\[-0.395cm]
  & Metrics & Summary measures of observed outcomes. & Conversion rates & 4 & \session{5,6,4,3} \\
  \\[-0.455cm]
\cmidrule(l{7.2em}){1-6}

\\[-0.4cm]

\bottomrule
\end{tabular}

\begin{tikzpicture}[remember picture, overlay]
  \draw[subsystem arrow] (exogenous.south) -- (reference.north);
  \draw[subsystem arrow] (reference.south) -- (dataprocessing.north);
  \draw[subsystem arrow] (dataprocessing.south) -- (serving.north);
  \draw[subsystem arrow] (serving.south) -- (application.north);
  \draw[subsystem arrow] (application.south) -- (response.north);
\end{tikzpicture}

\vspace{-1em}

\begin{tablenotes}
\small
\item[\textbf{Legend}] System elements are grouped in colored subsystem boxes ( \VarSymbol{externalColor!70} = Exogenous Influences; \VarSymbol{naturalColor!70} = Reference Domain; \VarSymbol{featureColor!70} = Processing Pipeline; \VarSymbol{inferenceColor!70} = Inference Pipeline; \VarSymbol{applicationColor!70} = Application; \VarSymbol{outcomeColor!70} = Endogenous Outcomes) and outlines indicate the system view ( \protect\dashedbox = environment; \protect\solidbox = ML system). (\FuncSymbol) illustrate relations between subsystems, ordered by runtime sequence: environment generates data, ML system processes and acts, producing outcomes.

\vspace{0.2em}

\item[*] This category extends the descriptive model of \cite{leest2025tea}; see Section \ref{suggestions} for suggested refinements to the model.
\end{tablenotes}
\vspace{0.5em}

\label{tab:categories-with-subsystem-column}

\end{threeparttable}  
\end{table*}

The resulting catalog of captured information is presented in Table~\ref{tab:categories-with-subsystem-column}, organized by system element and information type (as defined in Section \ref{background}). Table~\ref{tab:matrix} shows how this information supports specific monitoring activities. We structure this section around six monitoring activities that emerged from our analysis. For each activity, we address both research questions: (RQ1) what information practitioners systematically capture and how they use it, and (RQ2) what information remains uncaptured despite being recognized as critical for the activity.

Two elements emerge from practice that the descriptive model in \cite{leest2025tea} did not capture. Following our framework analysis approach (Section \ref{analysis}), we integrate these as extensions. In particular, \textit{Feature vectors} are a system element within inference pipeline representing the formatted inputs fed to models at inference time. \textit{Outcome variables} are a system element within the \textit{endogenous outcome} subsystem -- a new subsystem capturing observable outcomes (such as business metrics) generated after the ML system's actions influence the environment. We discuss these extensions to the descriptive model in Section \ref{suggestions}.

\newlength{\firstcolwidth}
\newlength{\secondcolwidth}
\newlength{\matrixwidth}

\setlength{\firstcolwidth}{0.8cm}
\setlength{\secondcolwidth}{6.4cm}
\setlength{\matrixwidth}{8.8cm}

\newlength{\headercolwidth}
\setlength{\headercolwidth}{\dimexpr\matrixwidth/6\relax}

\newlength{\matrixboxheight}
\setlength{\matrixboxheight}{0.34cm}

\newlength{\matrixrowheight}
\setlength{\matrixrowheight}{0.38cm}
\newlength{\matrixrowheightdouble}
\setlength{\matrixrowheightdouble}{0.38cm}

\newcommand{\VarSymbol}[1]{%
  \tikz[baseline=-0.6ex] \fill[#1, draw=none] (0,0) circle[radius=0.1cm];%
}

\definecolor{naturalColor}{RGB}{8,230,0}
\definecolor{LatentColor}{RGB}{69,150,66}
\definecolor{externalColor}{RGB}{164,224,34}
\definecolor{featureColor}{RGB}{71,179,255}
\definecolor{inferenceColor}{RGB}{56,112,232}
\definecolor{applicationColor}{RGB}{107,47,247}
\definecolor{outcomecolor}{RGB}{66,245,173}

\newcommand{\HeaderOne}{%
\begin{tikzpicture}[baseline=(current bounding box.center), remember picture]
  \node[
    fill=externalColor!40,
    draw=black,
    rounded corners=1pt,
    line width=0.6pt,
    minimum width=1.3cm,
    minimum height=0.2cm,
    anchor=center
  ] (H1) {};
\end{tikzpicture}%
}

\newcommand{\HeaderTwo}{%
\begin{tikzpicture}[baseline=(current bounding box.center), remember picture]
  \node[
    fill=naturalColor!40,
    draw=black,
    rounded corners=1pt,
    line width=0.6pt,
    minimum width=1.3cm,
    minimum height=0.2cm,
    anchor=center
  ] (H2) {};
\end{tikzpicture}%
}

\newcommand{\HeaderThree}{%
\begin{tikzpicture}[baseline=(current bounding box.center), remember picture]
  \node[
    fill=featureColor!40,
    draw=black,
    rounded corners=1pt,
    line width=0.6pt,
    minimum width=1.3cm,
    minimum height=0.2cm,
    anchor=center
  ] (H3) {};
\end{tikzpicture}%
}

\newcommand{\HeaderFour}{%
\begin{tikzpicture}[baseline=(current bounding box.center), remember picture]
  \node[
    fill=inferenceColor!40,
    draw=black,
    rounded corners=1pt,
    line width=0.6pt,
    minimum width=1.3cm,
    minimum height=0.2cm,
    anchor=center
  ] (H4) {};
\end{tikzpicture}%
}

\newcommand{\HeaderFive}{%
\begin{tikzpicture}[baseline=(current bounding box.center), remember picture]
  \node[
    fill=applicationColor!40,
    draw=black,
    rounded corners=1pt,
    line width=0.6pt,
    minimum width=1.3cm,
    minimum height=0.2cm,
    anchor=center
  ] (H5) {};
\end{tikzpicture}%
}

\newcommand{\HeaderSix}{%
\begin{tikzpicture}[baseline=(current bounding box.center), remember picture]
  \node[
    fill=outcomecolor!40,
    draw=black,
    rounded corners=1pt,
    line width=0.6pt,
    minimum width=1.3cm,
    minimum height=0.2cm,
    anchor=center
  ] (H6) {};
\end{tikzpicture}%
}

\newcommand{\MatrixBox}[4]{%
\begin{tikzpicture}[remember picture, overlay]
  \ifnum\pdfstrcmp{#3}{top}=0
    \coordinate (boxvertical) at ($(#4) + (0, 0.33cm)$);
  \else\ifnum\pdfstrcmp{#3}{bottom}=0
    \coordinate (boxvertical) at ($(#4) + (0, -0.0cm)$);
  \else\ifnum\pdfstrcmp{#3}{middle}=0
    \coordinate (boxvertical) at ($(#4) + (0, 0.04cm)$);
  \else
    \coordinate (boxvertical) at (#4);
  \fi\fi\fi
  \path let \p1 = (H#1.west), \p2 = (H#2.east), \p3 = (boxvertical) in
    node[
      draw=black!70,
      fill=black!5,
      line width=0.5pt,
      minimum height=\matrixboxheight,
      anchor=center,
      rounded corners=2pt,
      minimum width={veclen(\x2-\x1,0)}
    ] at ({(\x1+\x2)/2}, \y3) {};
\end{tikzpicture}%
}

\newcommand{\MatrixText}[5]{%
\begin{tikzpicture}[remember picture, overlay]
  \ifnum\pdfstrcmp{#3}{top}=0
    \coordinate (boxvertical) at ($(#4) + (0, 0.33cm)$);
  \else\ifnum\pdfstrcmp{#3}{bottom}=0
    \coordinate (boxvertical) at ($(#4) + (0, -0.0cm)$);
  \else\ifnum\pdfstrcmp{#3}{middle}=0
    \coordinate (boxvertical) at ($(#4) + (0, 0.04cm)$);
  \else
    \coordinate (boxvertical) at (#4);
  \fi\fi\fi
  \path let \p1 = (H#1.west), \p2 = (H#2.east), \p3 = (boxvertical) in
    node[
      draw=none,
      minimum height=\matrixboxheight,
      anchor=center,
      rounded corners=2pt,
      minimum width={veclen(\x2-\x1,0)}
    ] at ({(\x1+\x2)/2}, \y3) {\small #5};
\end{tikzpicture}%
}

\newcommand{\RowAnchorOne}{%
\raisebox{-0.5\height}{%
\raisebox{0.5\matrixrowheight}{%
\tikz[remember picture] \node[inner sep=0pt] (row1anchor) {};%
}}%
\rule{0pt}{\matrixrowheight}%
}

\newcommand{\RowAnchorTwo}{%
\raisebox{-0.5\height}{%
\raisebox{0.5\matrixrowheight}{%
\tikz[remember picture] \node[inner sep=0pt] (row2anchor) {};%
}}%
\rule{0pt}{\matrixrowheightdouble}%
}

\newcommand{\RowAnchorThree}{%
\raisebox{-0.5\height}{%
\raisebox{0.5\matrixrowheight}{%
\tikz[remember picture] \node[inner sep=0pt] (row3anchor) {};%
}}%
\rule{0pt}{\matrixrowheightdouble}%
}

\newcommand{\RowAnchorFour}{%
\raisebox{-0.5\height}{%
\raisebox{0.5\matrixrowheight}{%
\tikz[remember picture] \node[inner sep=0pt] (row4anchor) {};%
}}%
\rule{0pt}{\matrixrowheight}%
}

\newcommand{\RowAnchorFive}{%
\raisebox{-0.5\height}{%
\raisebox{0.5\matrixrowheight}{%
\tikz[remember picture] \node[inner sep=0pt] (row5anchor) {};%
}}%
\rule{0pt}{\matrixrowheight}%
}

\newcommand{\RowAnchorSix}{%
\raisebox{-0.5\height}{%
\raisebox{0.5\matrixrowheight}{%
\tikz[remember picture] \node[inner sep=0pt] (row6anchor) {};%
}}%
\rule{0pt}{\matrixrowheight}%
}

\begin{figure*}[!t]
\begingroup
\captionsetup{type=table} 

\centering
\normalsize
\setlength{\tabcolsep}{3pt}
\renewcommand{\arraystretch}{1.0}

\begin{threeparttable}
\caption{Results organized by monitoring activity, mapped to captured \textit{information} (gray) of the \textit{system context} (colored).}
\label{tab:matrix}

\begin{tabular}{m{\firstcolwidth}|m{\secondcolwidth}|p{\headercolwidth}p{\headercolwidth}p{\headercolwidth}p{\headercolwidth}p{\headercolwidth}p{\headercolwidth}}
\toprule

\multicolumn{2}{c|}{Section} &
\raisebox{0.12cm}{\HeaderOne} &
\raisebox{0.12cm}{\HeaderTwo} &
\raisebox{0.12cm}{\HeaderThree} &
\raisebox{0.12cm}{\HeaderFour} &
\raisebox{0.12cm}{\HeaderFive} &
\raisebox{0.12cm}{\HeaderSix} \\
\hline
\centering \S\ref{results_1} & \textbf{VALIDATE} Through Outcome Signals & \multicolumn{6}{c}{\RowAnchorOne} \\
\hline
\centering \S\ref{results_2} & \textbf{DETECT} Faults Through Instrumentation & \multicolumn{6}{c}{\RowAnchorTwo} \\
\hline
\centering \S\ref{results_3} & \textbf{LOCALIZE} a Fault Through DataFlows & \multicolumn{6}{c}{\RowAnchorFour} \\
\hline
\centering \S\ref{results_4} & \textbf{IDENTIFY} Root Causes With Provenance & \multicolumn{6}{c}{\RowAnchorThree} \\
\hline
\centering \S\ref{results_5} & \textbf{VERIFY} Correctness with Domain Knowledge & \multicolumn{6}{c}{\RowAnchorFive} \\
\hline
\centering \S\ref{results_6} & \textbf{INTERPRET} Shifts Through Domain Context & \multicolumn{6}{c}{\RowAnchorSix} \\
\bottomrule
\end{tabular}

\begin{tablenotes}
\small
\item[\textbf{Legend}] \VarSymbol{externalColor!70} \ = Exogenous Influences; \VarSymbol{naturalColor!70} \ = Reference Domain; \VarSymbol{featureColor!70} \ = Processing Pipeline; \VarSymbol{inferenceColor!70} \ = Inference Pipeline; \VarSymbol{applicationColor!70} \ = Application; \VarSymbol{outcomecolor!70} \ = Endogenous Outcomes.
\item[*] \textit{Attributes} apply to only a subset of subsystems, but are grouped for visual simplicity.
\end{tablenotes}
\end{threeparttable}

\MatrixBox{6}{6}{middle}{row1anchor}
\MatrixText{6}{6}{middle}{row1anchor}{Metrics}
\MatrixBox{3}{5}{middle}{row2anchor}
\MatrixText{3}{5}{middle}{row2anchor}{Metrics; Attributes*}
\MatrixBox{3}{3}{middle}{row4anchor}
\MatrixText{3}{3}{middle}{row4anchor}{Graphs}
\MatrixBox{3}{5}{middle}{row3anchor}
\MatrixText{3}{5}{middle}{row3anchor}{Metadata; Attributes*}
\MatrixBox{2}{2}{middle}{row5anchor}
\MatrixText{2}{2}{middle}{row5anchor}{Assertions}
\MatrixBox{1}{1}{middle}{row6anchor}
\MatrixText{1}{1}{middle}{row6anchor}{Identifiers}
\MatrixBox{2}{2}{middle}{row6anchor}
\MatrixText{2}{2}{middle}{row6anchor}{Slices}

\begin{tikzpicture}[remember picture, overlay]
\draw[-{Stealth[length=2mm]}, thick] (H1.east) -- (H2.west);
\draw[-{Stealth[length=2mm]}, thick] (H2.east) -- (H3.west);
\draw[-{Stealth[length=2mm]}, thick] (H3.east) -- (H4.west);
\draw[-{Stealth[length=2mm]}, thick] (H4.east) -- (H5.west);
\draw[-{Stealth[length=2mm]}, thick] (H5.east) -- (H6.west);
\end{tikzpicture}

\endgroup
\end{figure*}

\subsection{Validate through Outcome Signals}\label{results_1}
Practitioners track outcomes to validate whether the ML systems do have the desired impact. Yet, systematically capturing outcomes proves challenging and faces organizational barriers.

\subsubsection{Business metrics and experiments to estimate real-world impact}

Teams track \textbf{outcome metrics} to validate whether their ML systems achieve intended business impact (4/7 sessions; \session{3,4,5,6}) -- measuring customer satisfaction, conversion rates, or prevented losses that ultimately determine system success.

The most straightforward approach is tracking \textit{direct outcomes}, where teams track business metrics during model rollouts to test whether offline improvements translate to measurable gains \session{5,6}. But even this seemingly simple validation hits a wall. Attribution becomes nearly impossible because the internal system and external environment evolve continuously while effect sizes remain small. One practitioner captured the frustration \session{6}:

\begin{quote}
    \textit{``It's hard to see that back in online metrics... to attribute the change of a model... to the business KPI that are only observed all the way at the end.''}
\end{quote}

In some cases, a more fundamental measurement problem emerges. The system's own interventions censor the outcomes that the teams need to observe to validate the model \session{3,4,7}. Alternative outcomes effectively never occur and thus cannot be assessed -- e.g, content recommendations hide user responses to alternative content, or early interventions with at-risk patients prevent natural readmission patterns. Teams address this issue through two strategies. Some collect structured customer feedback from affected users \session{4}, asking directly whether interventions proved to be appropriate given the observed context. Others track \textit{potential outcomes} by deliberately withholding action from small holdout populations, enabling observation of what would happen without the model's intervention \session{3,5}. As one practitioner explained \session{3}:

\begin{quote}
    \textit{``We have control traffic, so there is a portion of the traffic that we do produce [predictions] for but we don't act on them.''}
\end{quote}

This interventional monitoring strategy actively elicits counterfactual outcomes that censoring would otherwise hide; the ability to observe outcomes thus requires controlled interventions that generate otherwise inaccessible signals.

\subsubsection{Barriers to systematic outcome capture}

Despite recognizing the need to track \textbf{outcome metrics}, many teams reported that systematically capturing \textbf{outcome metrics} remains challenging due to organizational constraints  (4/7 sessions; \session{1,6,7,4}). Some teams wanted to expand their capture of \textit{direct outcomes} but faced barriers. One team wanted to systematically capture customer feedback but lacked the necessary product integration; the capability simply does not exist in the user-facing interface \session{7}. Feedback delays further complicate validation, preventing teams from tracking validation results in reasonable timeframes \session{1,6}. Tracking \textit{potential outcomes} through holdouts faces another barrier. While teams recognized feedback censoring as a problem, regulatory constraints and business risk prevented them from withholding treatment \session{4,7}. The intervention needed to generate the monitoring signal violated compliance requirements or threatened unacceptable business losses. As one practitioner explained the asymmetric risk \session{4}:

\begin{quote}
\textit{``[False negatives] are costly... [false positives are] perceived as not as bad. But of course, it's better for us to have a high recall.''}
\end{quote}

Furthermore, continuous experimentation to systematically track (potential) \textit{outcome metrics} often loses out to other priorities. Teams at earlier maturity stages prioritized developing models and tracking model performance over establishing online experimentation \session{6}. Expensive experiments with small effect sizes reduced the priority level of outcome tracking -- teams chose velocity over validation infrastructure.

\subsection{Detect Faults through Instrumentation}\label{results_2}

All teams instrument their pipeline to emit metrics throughout the ML system -- from data ingestion through feature computation to model predictions -- aiming to detect failures close to where they originate. By tracking signals at each stage where faults might occur ``to make sure technical metrics catch things before the business metrics'' \session{6}. Yet as we discuss in Section \ref{detection_gap}, the reality proves more complex -- with gaps at system boundaries complicating fault detection.

\subsubsection{Signals captured across the internal system} \label{detection_captured}

Practitioners prioritize monitoring for internal system faults -- failures in data processing, model serving, and application logic. Among these, data integrity issues emerged as the dominant failure mode across sessions (6/7; \session{1,2,3,5,6,7}).



To catch these faults early, teams track data quality metrics across data processing stages. These metrics include a range of summary statistics: distribution divergence \session{2}, uniqueness \session{6}, min/max values \session{1,2,6}, and number of NULL values \session{1,2,3,4,5,6}. Teams reported a clear division of responsibility: tracking these metrics typically falls under data engineering teams rather than ML teams directly \session{1,2,3,6}, with one practitioner noting \session{6}: 

\begin{quote}
\textit{``The data warehouse team is testing whatever they provide to us as ML engineers.''}
\end{quote}

\textbf{Source table metrics} track data quality on raw data sources after ingestion, catching corrupted data transmitted from third parties or customers. Two sessions reported systematically tracking these metrics \session{2,6}, to catch problems ``early in the process'' \session{2}. 

\textbf{Feature table metrics} track the processed data that models actually consume, detecting when tables become stale (not updated) \session{1,6}, contain incorrect values \session{1,6}, or are not populated at all \session{1,3,6} -- any of which would degrade model outputs. Three sessions reported tracking data quality metrics over feature table columns \session{1,3,6}, serving as an early warning system before corrupted features reach production models. While these metrics target issues within feature tables themselves, they also surface upstream faults that cascade through table dependencies. As one practitioner described \session{6}:

\begin{quote}
    \textit{``The data didn't get refreshed. But that was a table that we based another table on... but we were just missing a part.''}
\end{quote}

Once data is processed through the data pipelines, ML teams add instrumentation to track the serving layer, as discussed below.

\textbf{Feature vector metrics} capture data quality at the moment of model inference, computing summary statistics like distribution divergence, uniqueness, max and min values, and NULL counts on the formatted inputs fed to the model (5/7 sessions; \session{2,3,5,6,7}). These signals catch local serving issues -- e.g., failed API calls to feature stores, or schema mismatches. However, practitioners predominantly tracked distribution metrics to monitor data state broadly -- capturing upstream faults and natural environmental changes propagating through the pipeline.

\textbf{Model attributes} capture operational measures of the serving infrastructure itself, including inference latency, throughput, and service health (3/7 sessions; \session{4,5,7}). Teams set alerts directly on these attributes, where drops in throughput or anomalies in volume automatically trigger alarms -- primarily intended to catch service failures in the serving infrastructure. Yet these signals also surface upstream faults cascading from an upstream source. As one practitioner discovered through a drop in model throughput \session{5}: 

\begin{quote}
    \textit{``We have a monitor for [throughput], and it decreased significantly... I discovered that somewhere upstream, one of the features wasn't sent to us, it was empty.''}
\end{quote}

\textbf{Prediction metrics} emerged as the primary signal ML teams actively monitor (6/7 sessions; \session{1,2,3,5,6,7}), serving as the main, catch-all indicator of upstream issues and model health. Most teams track aggregate metrics that capture changes in the distribution of predictions over time \session{1,2,3,5,6,7}, with some also actively tracking accuracy metrics when feedback is available -- a metric computed of prediction and outcome metrics jointly \session{3,4,6}.

Downstream from the model, teams track \textbf{action metrics} -- the decisions actually executed by application policies that consume model predictions (4/7 sessions; \session{1,3,4,7}). These outputs are produced by decision policies that take various forms -- rule-based systems where model scores inform conditions \session{1,3,6}, workflows that aggregate multiple model outputs \session{2,7}, or simple thresholds over model scores \session{4,5,7}. Practitioners report tracking both distribution divergence of executed actions \session{3,4,7} and the correctness of individual decision rules \session{1}. Teams set alerts on action metrics to catch local application faults: outdated thresholds that fail to adapt to model improvements \session{4,7}, or misconfigured rules that create technically valid but logically flawed decisions \session{3}. As one practitioner explained \session{3}:

\begin{quote}
\textit{``If the configuration doesn't make sense... or if there is a mistake, from technical point of view, it's a valid change, the system behaves normally, but logically the action wasn't correct.''}
\end{quote}

Yet as the final output point, action metrics effectively serve as a sink for issues originating anywhere in the system: data processing faults, model degradation, environmental shifts, and application logic errors all ultimately manifest in changed actions.

Tracking the correctness of actions involves cross-team collaboration, with business stakeholders who configure these policies actively monitoring alongside ML teams \session{1,3,6}. As one described, ``analysts are constantly eyeballing the time-series'' of the final decisions \session{3}. Practitioners viewed tracking actions and managing decision policies as integral to ML practice, with one noting \session{1}:

\begin{quote}
\textit{``However many models we have, we probably have two orders of magnitude more rules, and I view a rule as a simple machine learning model.''}
\end{quote}

\subsubsection{Missing instrumentation at system boundaries} \label{detection_gap}

Despite extensive instrumentation across the ML pipeline, practitioners reported that signals must comprehensively cover subsystems to be actionable -- otherwise gaps in coverage prevent effective triage (4/7 sessions; \session{1,2,3,6}). When downstream signals fire but upstream signals are missing, teams cannot determine the cause -- whether the alert indicates a critical failure or benign change. When upstream signals fire without corresponding downstream impact signals, teams cannot assess the impact -- whether the detected issue actually matters for system operation.

The previous subsection illustrated this challenge: prediction metrics and action metrics frequently surfaced issues originating in data processing or the environment, yet without comprehensive upstream instrumentation, practitioners faced manual investigation to understand what triggered the downstream alert. As one practitioner emphasized the need for complete boundary coverage rather than selective instrumentation \session{1}:

\begin{quote}
\textit{``We'd like to have [data quality tracked] at each interface, before going to the next.''}
\end{quote}

This gap proves especially challenging at organizational boundaries where different teams own different stages of the ML system. ML teams repeatedly encountered downstream anomalies caused by a fault in a subsystem managed by different teams -- including corrupted source data from third parties, stale feature tables from failed data engineering jobs, or outdated configurations \session{1,2,3,6}. 



However, broad instrumentation coverage presents its own challenges. Teams reported that extensive monitoring produces many false positives \session{3,6} -- detecting inconsequential changes that do not impact downstream predictions or actions, making it harder to identify critical issues. One team addresses this by composing signals and only treats an observed shift as operationally relevant when multiple signals shift jointly -- e.g., the NULL rate of a feature jointly with prediction divergence, as explained in \session{3}:

\begin{quote}
    \textit{``To get your signaling in order is pretty hard because you've lots of false positives, but if you combine a couple of signals, you get a much cleaner signal.''}
\end{quote}

\subsection{Localize a Fault Through DataFlows}\label{results_3}
When faults are detected, practitioners need to trace them back through their systems to identify where problems originated. This tracing depends on understanding how data and dependencies flow through the ML pipeline -- connecting downstream effects to upstream causes. Yet most teams lack systematic representations of these flows, instead relying on manual investigation through complex, undocumented systems.

\subsubsection{Graphical representations to systematically navigate dependencies}
A common challenge across teams was navigating complex dependencies -- especially pipelines with many intermediate steps and data loaded from different sources -- when tracing a fault to its root-cause (5/7; \session{1,2,3,5,6}). As one practitioner noted \session{1}:

\begin{quote}
\textit{``It's not obvious... how that final feature store came to be. There's all sorts of intermediate transformations which are dependencies on the final things. Many steps of transformations, many steps to create features, and features used to create other features.''}
\end{quote}

To address this, some teams maintain \textbf{transformation graphs} -- directed graphs showing how data flows from source tables through transformations to feature tables -- which they use to systematically trace faults back to their origins (2/7 sessions; \session{6,7}). When available, these graphs dramatically reduce diagnostic time. One team described quickly identifying which transformation caused a sample size drop \session{7}:

\begin{quote}
\textit{``It was obvious from the graph, we saw a decline in the samples we're using... we look at the individual filters we apply... and then we saw that's the culprit.''}
\end{quote}

The participants mentioned that these graphs are automatically derived using orchestration frameworks \session{6,7} such as dbt~\cite{dbt_labs_tool} and Luigi~\cite{spotify_luigi}.

\subsubsection{Missing structural overviews force manual fault tracing}
Despite the clear value of graphical representations, most teams reported lacking systematic visual overviews of their systems (5/7 sessions; \session{1,2,3,4,5}). Instead, they rely on mental models and manual investigation -- tracing system elements upstream from detected faults without structured guidance. This manual approach proves time-consuming and error-prone. As one participant described having \textit{``a pretty busy two weeks figuring out what happened''} \session{3}, tracking down a fault that originated from a configuration change in source data ingestion and propagated through multiple pipeline stages before surfacing as an increase in false positive predictions.

The challenge extends beyond data processing alone. Teams reported complex interdependencies spanning multiple system layers \session{1,2,3,7}: models feeding into other models \session{3,7}, decision policies aggregating outputs of multiple models \session{2,7}, and cascading chains where policies consume model outputs, feed decisions into subsequent models, which then inform further policies \session{3}. One practitioner captured the overwhelming scope of decision policies managed by different teams \session{3}:

\begin{quote}
\textit{``There are rules here, everywhere, rules before us, rules afterwards, rules between us... I think we're at the stage where everyone realizes that we cannot move anymore like this.''}
\end{quote}

While teams recognize the value of capturing structural overviews, it may not the first priority when immediate monitoring gaps demand attention \session{1}. Furthermore, connecting subsystems managed by different data, modeling, and infrastructure teams across organizational boundaries complicates the aspiration of building a unified dependency view spanning the entire ML system \session{3}.

\subsection{Identify Root Causes With Provenance}\label{results_4}

Upon receiving a fault signal, practitioners trace backwards through their systems to identify what changed upstream. Code deployments, configuration updates, or schema changes that coincide with detected anomalies become candidates for the root-cause. This backward tracing depends on provenance information -- the metadata and operational attributes that record what underlies a fault.

\subsubsection{Diagnostic information accessed during root-cause analysis}

Teams reported investigating faults by examining artifacts across different system layers. \textbf{Transformation metadata} captures the code and configurations that transform data through the pipeline \session{6,7}. Teams examine SQL scripts or Python transformations to identify logical faults -- filters that exclude too much data, joins with incorrect conditions, or transformations that introduce errors \session{6,7}. These errors often remain undetected for extended periods. As one practitioner discovered \session{6}:

\begin{quote}
\textit{``We did a join, and we did some filters, and I think after a few months, we noticed that one of our filters excluded a quarter of our [data].''}
\end{quote}

Teams capture metadata artifacts to trace faults back to their origins. \textbf{Source table metadata} is examined to identify schema issues \session{2,3}. \textbf{Model metadata} -- like training scripts -- were consulted to identify faults introduced during training that only surfaced in production \session{4,6}. \textbf{Policy metadata} is captured covering decision rule configurations enabling tracing faults to logical inconsistencies made by stakeholders \session{3}.

Beyond metadata, practitioners use operational measures to pinpoint fault location. \textbf{Transformation attributes} like throughput help identify which pipeline stages experienced issues \session{6,7}, while \textbf{feature table attributes} like staleness indicators help trace prediction faults to tables that failed to update \session{2}.

When tooling makes provenance information accessible within individual systems, investigation becomes straightforward. Databricks \cite{databricks_platform} exposed feature table staleness \session{2}, while Luigi \cite{spotify_luigi} logs captured throughput of individual operations -- enabling precise fault isolation \session{7}. A gap emerges when tooling doesn't automatically surface these connections in response to downstream alerts.

\subsubsection{Provenance disconnected from monitoring workflows}

Like most software systems, provenance information is often captured by default. Version information exists and is tracked, but automatic correlation with monitoring signals does not. As one practitioner explained \session{3}:

\begin{quote}
\textit{``You see the version, of course we're tracking it, but no one goes ahead and automatically correlates it, like there was a configuration before and then there's a configuration after, and immediately you see the change, this sharp distributional change.''}
\end{quote}

The gap is not that provenance metadata does not exist -- code is versioned, configurations are stored, schemas are tracked. The gap is that this metadata remains isolated from the monitoring signals that would make it actionable for root-cause analysis.

\subsection{Verify Correctness with Domain Knowledge}\label{results_5}

Without observing the ground truth, practitioners must rely on domain knowledge to evaluate whether the data processing produces correct data and their models produce correct outputs. This knowledge exists along a spectrum -- from formalized assertions that enable automated verification to tacit knowledge existing only in stakeholders' judgment.

\subsubsection{Supervision through domain assertions}

Some teams encode domain knowledge as explicit assertions -- logical rules that define valid ranges and expected behaviors for feature and target variables -- enabling automated detection of domain-violating outputs \session{2,6}. These assertions encode properties of the modeled domain -- the external real-world process -- rather than the internal system artifacts where they are applied.

\textbf{Feature assertions} test whether observed statistics align with known properties of feature variables in the modeled domain \session{2,6}. For example, an assertion that customer age must fall between 0 and 120 years encodes domain knowledge about real-world customers. This assertion serves as a verification rule against which observed metrics (such as min/max values) can be checked -- both in feature tables after data processing and in feature vectors at serving time; they verify that processed data conform to expected real-world patterns and are indeed correctly processed and served.

\textbf{Target assertions} monitor whether prediction frequencies align with known properties of the target variable in the modeled domain \session{2,6}. One team noted they leverage their understanding that the positive class rate should remain stable (e.g, at 5\%) to set alert thresholds \session{6}. When predictions suddenly jump (e.g., to 20\%), they recognize this as \textit{``a clear signal that either something is wrong with our population or something is wrong with the model''} \session{6}.

\subsubsection{Correctness criteria often remain tacit, existing only in stakeholder judgment}

Many teams rely on human-in-the-loop verification through ad hoc consultation or \textit{``an angry phone call''} \session{3,7}, leaving domain knowledge tacit rather than codified as systematic checks (5/7 sessions; \session{1,3,4,6,7}). Product teams and analysts informally monitor outputs, assessing correctness based on this domain expertise. As one practitioner noted \session{6}:

\begin{quote}
    \textit{``Our colleagues are our monitoring tool... whenever they see some unexpected output, we get notified.''}
\end{quote}

Some teams aspire to capture this tacit knowledge as more expressive assertions -- moving beyond simple range checks to conditional rules that encode expected model behavior in specific scenarios \session{4,6}. As one team described their plans \session{4}:

\begin{quote}
    \textit{``You have a bunch of scenarios and from domain knowledge, you know what you want from a model, how a model should behave... that really tests if the model that is currently in production doing what I expect from it.''}
\end{quote}

The challenge is to systematically elicit this knowledge from stakeholders and to determine which assertions and scenarios warrant formalization -- recognizing that exhaustive coverage of all possible cases remains fundamentally infeasible.

Beyond these practical challenges, correctness verification itself proved to be contested in all roles. Although domain expertise on target variables can be encoded as assertions to monitor correctness, practitioners reported tensions between technical and organizational definitions of correctness \session{6,7}. Data scientists often assess whether predictions align with population-level probabilities -- treating the target as a random variable with an underlying distribution \session{7}. If a model’s predictions reflect that distribution (e.g., a ~10\% base rate), they consider it correct. Business stakeholders, however, judge correctness against organizational requirements and expectations -- such as maintaining a desired rate of actions triggered by a threshold -- which may not align with the model’s predictions.

\subsection{Interpret Shifts Through Domain Context}\label{results_6}
Distribution shifts happen continuously, but distinguishing real problems from benign changes requires understanding the domain context. Practitioners slice metrics by customer segments to localize issues, but ultimately rely on manual consultation with analysts and domain experts to explain what events -- seasonality, campaigns, disruptions -- are driving the changes.

\subsubsection{Domain context to expose subgroup behavior}

Practitioners structure their monitoring around domain context by defining customer segments and tracking metrics separately for each subgroup -- using \textbf{exogenous identifiers} and \textbf{feature slices} to partition their data (4/7 sessions; \session{3,5,6,7}).

Teams structure their understanding of the domain through two forms of segmentation. \textbf{Exogenous identifiers} assign observations to business-defined categories such as geography, customer maturity, or market segment (3/7 sessions; \session{3,5,6}). These key-value fields partition observations by external context. \textbf{Feature slices}, by contrast, partition the feature space by behavioral patterns within the data (1/7 sessions; \session{7}). For example, ``customer age'' might be sliced into young (0-18), adult (18-65), and senior (65+) subgroups.

Both enable segment-level monitoring of instrumented metrics or business outcomes across countries, customer cohorts, or product categories \session{6,5,3,7}. This contextual understanding helps explain metric variations; different groups exhibit distinct behaviors and respond differently to external events in the environment -- distribution shifts are often localized, requiring monitoring that captures these differences across subgroups. For example, a national holiday might affect sales in one country but not others; a sales campaign might impact new customers differently than established ones.

By partitioning data streams into semantically consistent groups, practitioners can explain issues that would otherwise be obscured in aggregate metrics. One practitioner explained the first step after detecting a shift in predictions \session{3}:

\begin{quote}
    \textit{``The output probabilities shifted as well. For some [subgroups], not so much for some... more false positives, and for others [more false negatives]. So there was like a bit of a mix of problems across [subgroups].''}
\end{quote}

The fact that all subgroups exhibited changes -- though with different symptoms -- provided an early indication of an internal fault. Indeed, it turned out to be an outage of an upstream dependency.

\subsubsection{Reliance on a human-in-the-loop instead of systematic contextualization}

Practitioners easily detect distribution shifts but struggle to distinguish technical failures from benign distribution shifts \session{1,2,3,4,6}. As one practitioner described this ambiguity \session{3}:

\begin{quote}
    \textit{``It's not so hard to monitor when [metrics] go up or down. No. Tricky thing is, they naturally go up or down.''}
\end{quote}

Teams actively consult domain experts -- typically analysts with real-time understanding of the domain -- to explain why distributions shift \session{1,6}. These experts provide context about trends or events -- like, marketing campaigns, or supply chain disruptions that \textit{``get flagged by a business analyst''} \session{6} -- that only become visible as distribution shifts in the observed data. As one practitioner explained their dependency on analysts \session{1}:

\begin{quote}
    \textit{``Because [analysts] see the data live, they also pick up on new trends quickly. And we have mechanisms in place [to communicate this] back to us.''}
\end{quote}

This reliance on human judgment creates two critical challenges. First, without systematic capture of external context, teams cannot automate their response -- each alert requires manual consultation \session{1,6}. Second, without understanding whether shifts reflect transient or persistent changes, automated retraining risks overfitting to temporary patterns. As one practitioner described \session{3}:

\begin{quote}
    \textit{``What we often have is like [a transient event that] impacts our training data quite a bit, so the model is retrained, adapts to it, but then during inference, this pattern disappears. So performance goes down, because we're overfitting to some non causal pattern.''}
\end{quote}

Teams reported they lacked mechanisms to record this information; they viewed the external environment as \textit{"a complex black-box system"} \session{3} with no means to track events in real-time \session{3,4}. One team recognized this limitation and aspired to capture the exogenous context in their monitoring workflow \session{7}:

\begin{quote}
    \textit{``I think in our monitoring we need to be aware on social cues and social news that would affect and help us explain to our stakeholders, why are the [metrics] high during those periods.''}
\end{quote}



\section{Discussion}\label{discussion}

We characterized ML observability practices across organisations in various domains. Below, we answer the two research questions defined in our study design (Section \ref{study_design}) referring to current observability practices (RQ1) and identified gaps in observability (RQ2).

\textbf{Answer to RQ1.} Teams consistently capture information about both the ML system and external environment. They instrument data processing, model serving, and application layers to detect faults, and track outcome metrics when feasible. For fault investigation, teams capture artefact metadata and attributes for tracing, with some maintaining dependency graphs for systematic root-cause analysis. Teams represent domain understanding by defining meaningful subgroups to read signals in context and by implementing rule-based checks to confirm correctness.

\textbf{Answer to RQ2.} Instrumentation to detect faults remains incomplete, particularly across elements owned by different teams, while outcome metrics are often infeasible due to organizational constraints. With few documented dependencies beyond those inferred by orchestration, and provenance metadata detached from monitoring, root-cause analysis remains mostly manual. Interpretation depends predominantly on tacit knowledge -- including external events that explain distribution shifts and domain expertise to verify correctness -- which is often not systematically captured and therefore relies heavily on human-in-the-loop practices.

Taken together, these results motivate refinements to the descriptive model (Section~\ref{suggestions}) and implications for practice (Section~\ref{implications}); we also note limitations (Section~\ref{limitations}), related work (Section~\ref{related_work}), and future work (Section~\ref{future_work}).

\subsection{Reflection on Descriptive Model} \label{suggestions}

The literature-derived model \cite{leest2025tea} effectively categorized most practitioners' monitoring information (Section \ref{background}), demonstrating adequate coverage and discriminant validity. Recurrent unmappable content showed systematic gaps in the original model. Following our framework analysis methodology (Section \ref{analysis}), we propose the refinements below based on the empirical results of our study.

\textbf{Consolidate to simplified information types.} As noted in Section \ref{information_types}, the descriptive model characterizes information through aspect (states, structures, or properties) and representation (formal or informal) dimensions. While these can combine in different ways -- e.g., properties of a system element could be captured as logical assertions, probabilistic models, or textual descriptions -- practitioners consistently used a constrained set of combinations. We recommend adopting these simplified categories (identifiers, slices, assertions, metadata, attributes, metrics, graphs) for operationalizing the model in practice -- e.g., as a domain-specific language (DSL) -- while preserving the full framework for academic purposes.

\textbf{Add an endogenous outcome subsystem.} As reported in Section \ref{results_1}, teams extensively track business outcomes and impact metrics -- but the current descriptive model cannot represent the feedback loops where ML system actions influence the environment and generate these outcomes. We propose adding an endogenous outcome subsystem -- as defined in Section \ref{results} -- within the environment view to capture how actions generate observable outcomes.

\textbf{Add a feature vector system element.} As discussed in Section \ref{results_2}, teams systematically monitor feature vectors at serving time separately from stored feature tables. The distinction matters operationally -- serving-infrastructure failures surfaced only in vectors, not tables -- and it mirrors typical ownership (ML engineers monitor served features; data engineers monitor feature tables. We propose adding a separate feature vector element to the inference pipeline subsystem, while representing produced features as a feature table element in the processing pipeline, for consistency with practice.

\textbf{Remove the latent influences subsystem.} When explaining the \textit{system context} dimension, we noted persistent ambiguity between \textit{latent} and \textit{exogenous} \textit{influences}. Both describe external drivers in the environment, and their only difference -- being measured directly or inferred -- is already handled by the \textit{representation} dimension. Maintaining a separate subsystem therefore adds redundant complexity without improving discriminative value.

\subsection{Implications} \label{implications}

Our results have some critical implications for both tool development and future research directions, which we discuss here.

\subsubsection{Implication for practice and tooling support}

Current ML monitoring tools address different subsets of the system information catalogued in Table \ref{tab:categories-with-subsystem-column}, creating fragmentation that likely contributes to the gaps we observed. ML observability platforms (Arize \cite{arize}, Fiddler \cite{fiddler}, WhyLabs \cite{whylabs}) capture model-centric signals -- prediction metrics, feature statistics, and sliced performance analysis. Data observability platforms (Monte Carlo \cite{montecarlo}, Datadog \cite{datadog}) capture pipeline information -- table metrics, job attributes, and auto-derived dependency graphs for fault tracing. Cloud ML services (Vertex AI \cite{vertexai}, SageMaker \cite{sagemaker}) provide integrated serving metrics and drift detection over model inputs and predictions. Assertion frameworks (Great Expectations \cite{greatexpectations}, deequ \cite{deequ}) enable encoding domain constraints for automated verification across the pipeline.

This fragmentation likely contributes to the reported gaps in this study (Section \ref{results}). Diagnosing issues requires instrumentation and tracing across subsystems, yet tool support remains narrow. As argued in \cite{leest2025tea}, the descriptive model could provide a conceptual foundation for modern ML monitoring tooling by organizing information in a graph structure connecting information to system elements and elements to each other. This study defines the scope ML monitoring tools need to cover for completeness: systematically capturing information from the processing pipeline through the inference pipeline to downstream applications and the external environment -- far broader than current tools address.

\subsubsection{Implications for research}

The descriptive model synthesized from prior literature places significant emphasis on capturing information about the external environment \cite{leest2025tea}. Practice shows limited reliance on this information and a narrower range of information types than the literature emphasizes. This does not invalidate the theoretical framework or prior research, but it shows that proposed capabilities remain largely aspirational. Understanding this gap offers important directions for future research.

For example, prior research has developed sophisticated methods for capturing domain knowledge. Some approaches model scenarios probabilistically, describing how known external events affect the data-generating process to enable proactive detection \cite{sobolewskip2017scr,leestj2024expert}. Others propose textual playbooks and documentation frameworks that capture scenarios of external events and failure modes to support monitoring governance \cite{apaul2024mlops,dreyfuspa2022databased,feng2022clinical,haidert2021domain}. Others advocate utilizing causal graphs of the feature and target variables -- capturing the structure of the data-generating process, enabling systematic attribution of distribution shifts \cite{fengj2024designing,zhangh2023why,budhathokik2021why}.

None of our participants implemented these capabilities in production, despite their demonstrated theoretical merit, suggesting a gap between research capabilities and practical adoption.



\subsection{Limitations}\label{limitations}

Our study design involves several inherent constraints. First, focus groups capture reported practices rather than observing monitoring practice as conducted -- some tacit knowledge or ad hoc practices may be underrepresented. Second, we did not triangulate focus group data with analysis of actual monitoring artifacts such as dashboards, alert configurations, or incident tickets -- examining these artifacts might reveal practices not articulated during sessions. Finally, resource constraints limited engagement depth with each team -- extended case studies might provide a more complete characterization of observability not surfaced in our sessions.

\subsection{Related Work}\label{related_work}

Software observability practices are well-established -- interview studies confirm traditional monitoring frameworks effectively support practitioners despite remaining challenges \cite{niedermaier2019observability,li2022enjoy,davidson2023qualitative}. ML observability, however, is less mature, with exploratory and visionary attempting to establish the practice \cite{nogare2024machine,palumbo2023observability,shankars2022towards}.

Several empirical studies have examined ML monitoring from different perspectives. Interview studies in the broader MLOps space consistently surface monitoring as a significant challenge \cite{shankars2024we,muirurid2022practices,amrit2025analysis}, while more focused interview work specifically examines ML monitoring practices \cite{shergadwalamn2022a,protschky2025gets}. In similar, work, a questionnaire-based study confirms that tool support remains limited and no established framework for ML observability has emerged \cite{zimelewicz2024ml}. Multivocal literature reviews synthesizing practitioner discussions identify key guidelines and challenges, including the need to track comprehensive metrics and understand the causes behind observed distributional changes \cite{naveed2025monitoring,protschky2025gets}. Related surveys on trustworthy AI highlight monitoring challenges around fairness and explainability \cite{bayram2025towards,chen2024fairness}.

Experience reports and case studies shaped most of our understanding of ML monitoring and observability practice. Initial experiences from practice identified monitoring and observability challenges unique to ML that informed subsequent research \cite{sculley2015hidden}. Case studies document operational ML systems \cite{amershi2019software,ackermann2018deploying}, highlighting both challenges and workarounds practitioners employ. Maturity assessments show that less mature organizations rely predominantly on manual monitoring, with automated practices emerging as organizations advance \cite{john2021towards}. Case studies specific to ML monitoring yielded results on particular aspects like continuous data validation \cite{lelwakatare2021on}, and drift detection frameworks in practice \cite{xuz2023alertiger,zhoux2019a,mallick2022matchmaker}.

Our work addresses a  gap in this body of research. Although prior studies have classified monitoring challenges and documented practitioner pain points \cite{shankars2024we,muirurid2022practices,shergadwalamn2022a,naveed2025monitoring,protschky2025gets,zimelewicz2024ml}, none has systematically characterized what information practitioners capture for ML monitoring. We provide the first holistic empirical characterization of ML observability practices -- beyond documenting isolated challenges and best practices.


\subsection{Future Work}\label{future_work}

Our findings need validation across broader contexts. This study provides an initial empirical characterization of ML observability practice through focus groups across organizations. In-depth case studies are needed to validate and extend our findings -- particularly to examine the cross-role nature of ML monitoring we observed. Our results showed involvement across stakeholders: data engineers tracking pipeline quality, ML teams monitoring predictions, analysts verifying decisions, and domain experts interpreting shifts. How these roles monitor ML systems and capture information to enable observability warrants future research.

Beyond validation, future research should characterize patterns of system context information needed to detect and diagnose specific failure modes. While we catalogued what practitioners capture broadly, understanding which information combinations enable effective detection of particular faults -- data corruption, model degradation, policy misconfiguration -- remains unknown. Empirical studies could identify these patterns and evaluate whether implementing the corresponding capture mechanisms (automated provenance correlation, dependency graphs, formalized domain knowledge) improves monitoring effectiveness in practice.

To translate validated practices into practical tooling, an avenue for future work is to develop observability tooling that operationalizes the information surfaced here (e.g., through a DSL \cite{kourouklidis2023domain}).

\section{Conclusion} \label{conclusion}

This study presents the first empirical characterization of ML observability practices through focus groups with 37 practitioners in 7 organizations. We classified the information teams systematically capture across both the ML system -- signals spanning processing pipelines, inference pipelines, and applications -- and the external environment -- customer segments, domain constraints, and outcome metrics. Our results validate and refine a descriptive model for ML observability, confirming that effective monitoring requires capturing information about both system views. As such, we contribute a sound basis for developing practical ML observability.



\end{document}